\documentclass[11pt,3p]{elsarticle}

\usepackage{graphicx}
\usepackage{amsmath}
\usepackage{amssymb}
\usepackage{url}
\usepackage{hyperref}
\usepackage[english]{babel}
\usepackage{color}

\newcommand{\pT}{p_{_{\rm T}}}
\newcommand{\sqrts}{\sqrt{s}}
\newcommand{\pp}{p-p}
\newcommand{\ppbar}{p-$\overline{\rm p}$}
\newcommand{\epem}{{\rm e^+e^-}}

\providecommand{\mean}[1]{\ensuremath{\left<#1\right>}}

\journal{Nuclear Physics B}

\begin{document}

\begin{frontmatter}

\title{Confronting current NLO parton fragmentation functions with inclusive charged-particle spectra at hadron colliders}
\author{David d'Enterria}
\ead{dde@cern.ch}
\address{CERN, PH Department, CH-1211 Geneva 23, Switzerland}
\author{Kari J. Eskola}
\ead{kari.eskola@jyu.fi}
\author{Ilkka Helenius}
\ead{ilkka.helenius@jyu.fi}
\author{Hannu Paukkunen}
\ead{hannu.paukkunen@jyu.fi}
\address{Department of Physics, University of Jyv\"askyl\"a, P.O. Box 35, FI-40014, Finland}
\address{Helsinki Institute of Physics, University of Helsinki, P.O. Box 64, FI-00014, Finland}

\begin{abstract}

The inclusive spectra of charged particles measured at high transverse momenta ($\pT\gtrsim$~2~GeV/c) in
proton-proton and proton-antiproton collisions in the range of center-of-mass energies
$\sqrts$~=~200--7000~GeV  are compared with next-to-leading order perturbative QCD calculations using
seven recent sets of parton-to-hadron fragmentation functions (FFs). 
Accounting for the uncertainties in the scale choices and in the parton distribution functions,
we find that most of the theoretical predictions tend to overpredict the
measured LHC and Tevatron cross sections by up to a factor of two.  
We identify the currently too-hard gluon-to-hadron FFs
as the probable source of the problem, and justify the need to refit the FFs using the available LHC
and Tevatron data in a region of transverse momenta, $\pT\gtrsim$~10~GeV/c, 
which is supposedly free from additional non-perturbative contributions and 
where the scale uncertainty is only modest.

\end{abstract}


\end{frontmatter}

\section{Introduction}

The inclusive production of large-transverse-momentum ($\pT\gtrsim$~2~GeV/c) hadrons at proton-proton (\pp)
and proton-antiproton (\ppbar) colliders provides a ground for testing the factorization theorem of Quantum
Chromodynamics (QCD)~\cite{Collins:1989gx,Collins:1985ue} that predicts the universality and evolution of the two
non-perturbative elements in the theoretical cross-sections: parton distribution
functions (PDFs) and parton-to-hadron fragmentation functions (FFs). While the PDFs can nowadays be
tested and constrained by a multitude of different processes in deep-inelastic scattering and hadronic
collisions~\cite{Forte:2013wc}, the variety and kinematic reach
of data by which the FFs can be determined is more limited~\cite{Albino:2008aa,Albino:2008gy}. 
In this context, the inclusive hadron measurements at the LHC extending to unprecedentedly large values of center-of-mass energy
($\sqrt{s}$) and hadron $\pT$~\cite{Sassot:2010bh}, are particularly useful for
studying the FFs and their universality. From an experimental perspective, the best
precision and widest kinematic reach in the $\pT$-differential cross sections are achieved when no particle
identification is performed. The pre-LHC measurements for unidentified charged-hadron
spectra in \pp\ and \ppbar\ collisions range from fixed-target experiments at
$\sqrt{s}~\lesssim~60 \, {\rm GeV}$~\cite{Akesson:1982sc,Breakstone:1995ug}
to collider experiments covering a wide range of center-of-mass energies $\sqrt{s} = 200 - 1960 \, {\rm GeV}$
\cite{Adler:2005in,Adams:2006nd,Arsene:2004ux,Bocquet:1995jr,Albajar:1989an,Arnison:1982ed,Abe:1988yu,Aaltonen:2009ne}.
However, most of these data are restricted to moderate values of $\pT \lesssim  10 \, {\rm GeV/c}$, 
and the accuracy for $\pT \geq  10 \, {\rm GeV/c}$ is rather poor.
The importance of the new LHC data to overcome such limitations is underscored by the confusion
\cite{Arleo:2010kw,Albino:2010em,Cacciari:2010yd} triggered by the original CDF data~\cite{Aaltonen:2009ne}
which seemed to display deviations from next-to-leading order (NLO) perturbative QCD (pQCD) 
calculations up to three orders of magnitude at $\pT \simeq 150 \, {\rm GeV/c}$, but which was later on identified as
an experimental issue (see the erratum of Ref.~\cite{Aaltonen:2009ne}).\\ 

Interestingly, the NLO pQCD predictions presented along with the recently
published CMS~\cite{Chatrchyan:2011av,CMS:2012aa} and ALICE~\cite{Abelev:2013ala} inclusive charged hadron
spectra, appear to overshoot the data by up to a factor of two in the kinematical region where effects such as e.g. 
intrinsic transverse momentum of the colliding partons (intrinsic $k_T$)~\cite{Apanasevich:1998ki,Owens:1986mp,Wang:1998ww},
soft-gluon resummation~\cite{deFlorian:2008wt,deFlorian:2005yj}, or small-$z$ instabilities 
of the FFs~\cite{deFlorian:1997zj} should not play a major role. Especially since the recent LHC
measurements of the $\pT$-differential cross sections for inclusive
jets~\cite{Chatrchyan:2012bja,Aad:2011fc} and prompt photons~\cite{Chatrchyan:2011ue,Aad:2011tw} are
in perfect agreement with the NLO pQCD expectations~\cite{Wobisch:2011ij,Nagy:2003tz,d'Enterria:2012yj}, 
the data-vs-theory discrepancies for inclusive charged-hadrons come totally unexpected.
Resolving such inconsistencies is also of relevance for other QCD
analyses such as those related to the suppression of high-$\pT$ hadrons in
ultrarelativistic heavy-ion collisions where the \pp\ spectra are required as baseline
measurements~\cite{d'Enterria:2010zz}.\\

In this paper, we present a systematic comparison of the theoretical predictions for unidentified charged-hadron production
to experimental data with a special emphasis on the latest LHC measurements. Our aim is to
demonstrate that such a process in hadron colliders is predominantly sensitive to the gluon-to-hadron FFs which are presently
not well determined and, consequently, large discrepancies among the modern sets of FFs exist. These differences not only translate
into a significant scatter in the corresponding predictions for the cross sections, but 
none of the current FF sets can consistently reproduce the current LHC and Tevatron data at $\pT\gtrsim$~10~GeV/c. 
As the data measured by different experiments at the same collision energies are in mutual agreement, it seems excluded that such discrepancies are due to an experimental issue.  Instead, this hints to a severe problem in
the gluon-to-hadron FFs in most of the existing sets. 
We conclude that the gluon FF, which is currently mildly constrained by charged-hadron
spectra from hadronic collisions at RHIC and Sp$\overline{\rm p}$S energies, should be refitted by using the LHC
and Tevatron hadron spectra in the region $\pT\gtrsim$~10~GeV/c,
where the theoretical scale uncertainties appear tolerable and which should be
free from additional, non-perturbative hadron production mechanisms.

\section{The pQCD framework for inclusive hadron production}

The cross section for the inclusive production of a single hadron $h_3$ with a momentum $p_3$ 
in the collision of two hadrons $h_1$ and $h_2$ carrying momenta $p_1$ and $p_2$ respectively,
can be expressed, differentially in transverse momentum $\pT$ and (pseudo)rapidity $\eta$, as~\cite{Owens:1986mp,Aversa:1988vb}
\begin{eqnarray}
& & \frac{d\sigma(h_1 + h_2 \rightarrow h_3 + X)}{d\pT d\eta}  = 
 \sum _{ijl} \int dx_1 \int dx_2 \int \frac{dz}{z} f_i^{h_1}(x_1,\mu^2_{\rm fact}) f_j^{h_2}(x_2,\mu^2_{\rm fact}) \nonumber\\
 & & \hspace{-0cm} D_{l \rightarrow h_3}(z,\mu^2_{\rm frag}) \, \frac{d\hat{\sigma}(\hat p_1^i + \hat p_2^j \rightarrow \hat p_3^l, \mu^2_{\rm ren}, \mu^2_{\rm fact}, \mu^2_{\rm frag})}{d\hat p_{3T} d\eta}
_{\bigg|
{{{ \scriptstyle \hat p_1 = x_1 p_1} \atop {\scriptstyle  \hat p_2 = x_2 p_2}} \atop {\scriptstyle  \hat p_3 = p_3/z} }}
\label{eq:facinH} .
\end{eqnarray}
In this expression,
$f_i^{h_k}(x_k,\mu^2_{\rm fact})$
denote the PDFs of the
colliding hadrons evaluated at parton fractional momenta
$x_k$
and scale $\mu^2_{\rm fact}$. 
We will use the \textsc{ct10nlo}~\cite{Lai:2010vv} parametrization throughout this work.
The parton-to-hadron FFs are denoted by $D_{l \rightarrow h_3}(z,\mu^2_{\rm frag})$ where $z$ is the fraction
of the parton momentum carried out by the outgoing charged hadron. The PDFs and FFs are convoluted with
the partonic coefficient functions $d \hat\sigma$ for which we use their fixed-order NLO $\mathcal{O}(\alpha_s^3)$ 
expressions~\cite{Aversa:1988vb,deFlorian:2002az,Jager:2002xm}, treating the partons and hadrons as massless
particles. In practice, we evaluate these cross sections employing the \texttt{INCNLO}~\cite{Aversa:1988vb, INCNLO} program
which we have modified to improve the convergence at small values of $\pT$. The fixed-order calculations
are supposed to be adequate for $\pT \gg 1 \, {\rm GeV/c}$ but still sufficiently away from the phase-space
boundary $\pT^{\rm max} \sim \sqrt{s}/2$ (at midrapidity, $\eta\approx$~0), where
soft-gluon resummations~\cite{deFlorian:2008wt,deFlorian:2005yj} become relevant due to large logarithmic
contributions from an incomplete cancellation of the infrared divergences.\\

Truncating the partonic coefficient functions to $\mathcal{O}(\alpha_s^3)$ leads to the well-known
scale dependence of the pQCD calculations. For inclusive hadron production, there are three independent scales:
the renormalization scale $\mu_{\rm ren}$, factorization scale $\mu_{\rm fact}$, and the fragmentation
scale $\mu_{\rm frag}$. The sensitivity of the computed cross sections to the variations of these scales
is typically taken as an indication of the size of the missing higher-order corrections. Our default choice
is $\mu_{\rm ren} = \mu_{\rm fact} = \mu_{\rm frag} = \pT$, and
we take the scale uncertainty as the envelope enclosed by the following
16 scale variations~\cite{Banfi:2010xy}
\begin{equation}
\left( \frac{\mu_{\rm fact}}{\pT}, \frac{\mu_{\rm ren}}{\pT}, \frac{\mu_{\rm frag}}{\pT} \right) =
\begin{array}{cccc}
({1 \over 2},{1 \over 2},{1 \over 2}), & ({1 \over 2},{1 \over 2},1), & ({1 \over 2}, 1,{1 \over 2}), & ({1 \over 2},1,1), \\
({1 \over 2},1,2)                    , & (1,{1 \over 2},{1 \over 2}), & (1,{1 \over 2},1)           , & (1,1,{1 \over 2}), \\
(1,1,2)                              , & (1,2,1)                    , & (1,2,2)                     , & (2,1,{1 \over 2}), \\
(2,1,1)                              , & (2,1,2)                    , & (2,2,1)                     , & (2,2,2). 
\end{array}
\label{eq:scalevariation}
\end{equation}
We omit the combinations in which $\mu_{\rm ren}$ and $\mu_{\rm fact}$ or $\mu_{\rm ren}$ and $\mu_{\rm frag}$
are pairwise scaled by a factor of two in opposite directions due to the appearance of potentially large contributions of
the form $\log(\mu^2_{\rm ren}/\mu^2_{\rm fact})$ and $\log(\mu^2_{\rm ren}/\mu^2_{\rm frag})$ in the calculation.
The next-to-NLO (NNLO) calculations are expected to definitely reduce the scale dependence. However, although the PDF
analyses can be nowadays carried out partly at NNLO level~\cite{Ball:2012cx,Martin:2009iq,Gao:2013xoa,Alekhin:2013nda},
the time-like splitting functions needed in the NNLO evolution of the FFs are not yet fully known~\cite{Mitov:2006ic,Almasy:2011eq},
nor are the NNLO coefficient functions needed in Eq.~(\ref{eq:facinH}), although the latter could finally emerge
through the work currently carried out for jets~\cite{Ridder:2013mf,Currie:2013dwa,deFlorian:2013qia}. 

\section{Comparison of the parton fragmentation functions}

Table~\ref{tab:FFs} lists the seven commonly used sets of NLO parton-to-charged-hadron FF parametrizations: 
Kretzer (\textsc{kre})~\cite{Kretzer:2000yf}, \textsc{kkp}~\cite{Kniehl:2000fe},
\textsc{bfgw}~\cite{Bourhis:2000gs}, \textsc{hkns}~\cite{Hirai:2007cx}, \textsc{akk05}~\cite{Albino:2005me}
\textsc{dss}~\cite{deFlorian:2007aj,deFlorian:2007hc}, and \textsc{akk08}~\cite{Albino:2008fy}, that we employ
in our calculations. In a few of these analyses the charged hadron FFs are constructed as a sum of the individual FFs for pions
($\pi^\pm$), kaons ($K^\pm$), plus (anti)protons; 
but e.g. in \textsc{dss} there is still a small ``residual'' contribution on top of these. There are
other sets of FFs with restricted particle species~\cite{Binnewies:1994ju,deFlorian:2007aj,Soleymaninia:2013cxa}, 
but we do not consider them here as we focus only on the inclusive sum which should be better constrained than
the FFs for individual hadron species. For reviews, see e.g.~\cite{Albino:2008aa,Albino:2008gy}.\\

\begin{table}[htbp!]
\caption{Characteristics of the existing sets of parton-to-charged-hadron FFs. The hadron species included,
use of different collision systems, attempts to estimate the FF errors, minimum value of $z$ considered, and
the available $Q^2$-range are indicated.}
\begin{footnotesize}
\begin{center}
\begin{tabular}{lccccc}\hline
FF set  & Species & Fitted data & Error estimates & $z_{{\rm min}}$ & $Q^2$ (GeV$^2$)\\ \hline
Kretzer (\textsc{kre}) \cite{Kretzer:2000yf} & $\pi^\pm$,\,$K^\pm$,\,$h^+$+$h^-$ & $\epem$ & no & 0.01  & 0.8--$10^6$\\ 
\textsc{kkp} \cite{Kniehl:2000fe} & $\pi^+$+$\pi^-$,\,$K^+$+$K^-$ & $\epem$ &  no & 0.1  & 1 --$10^4$ \\
 & $p+\bar{p}$,\,$h^+$+$h^-$ &  &   &   &  \\
\textsc{bfgw} \cite{Bourhis:2000gs} & $h^\pm$ & $\epem$ & yes & $10^{-3}$  & 2--1.2 $\cdot 10^4$\\
\textsc{akk05} \cite{Albino:2005me}  & $\pi^\pm$,\,$K^\pm$,\,$p$,\,$\bar{p}$ & $\epem$ &  no & 0.1  & 2--$4 \cdot 10^4$\\
\textsc{hkns} \cite{Hirai:2007cx} & $\pi^\pm$,\,$K^\pm$,\,$p+\bar{p}$ & $\epem$ &  yes & $0.01$ & $1$--$10^8$ \\
\textsc{akk08} \cite{Albino:2008fy} & $\pi^\pm$,\,$K^\pm$,\,$p$,\,$\bar{p}$ & $\epem$,\,\pp  &  no & 0.05  & 2--$4 \cdot 10^4$\\
\textsc{dss} \cite{deFlorian:2007aj,deFlorian:2007hc} & $\pi^\pm$,\,$K^\pm$,\,$p$,\,$\bar{p}$,\,$h^\pm$ & $\epem$,\,\pp,\,e-p  &  yes & 0.05 & 1--$10^5$ \\ 
\hline
\end{tabular}
\end{center}
\end{footnotesize}
\label{tab:FFs}
\end{table}

\begin{figure}[tbhp]
\centering
\includegraphics[width=0.49\textwidth]{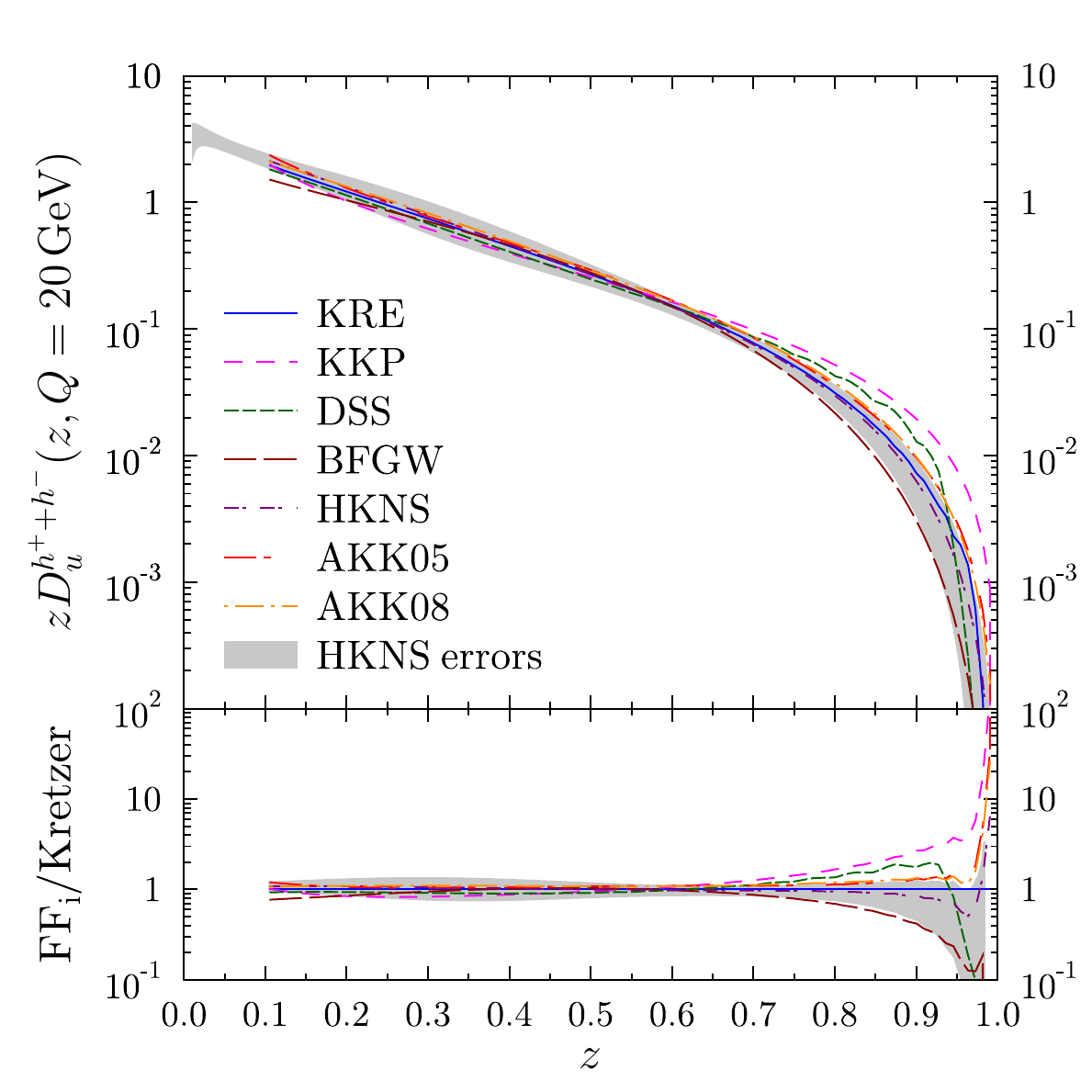}
\includegraphics[width=0.49\textwidth]{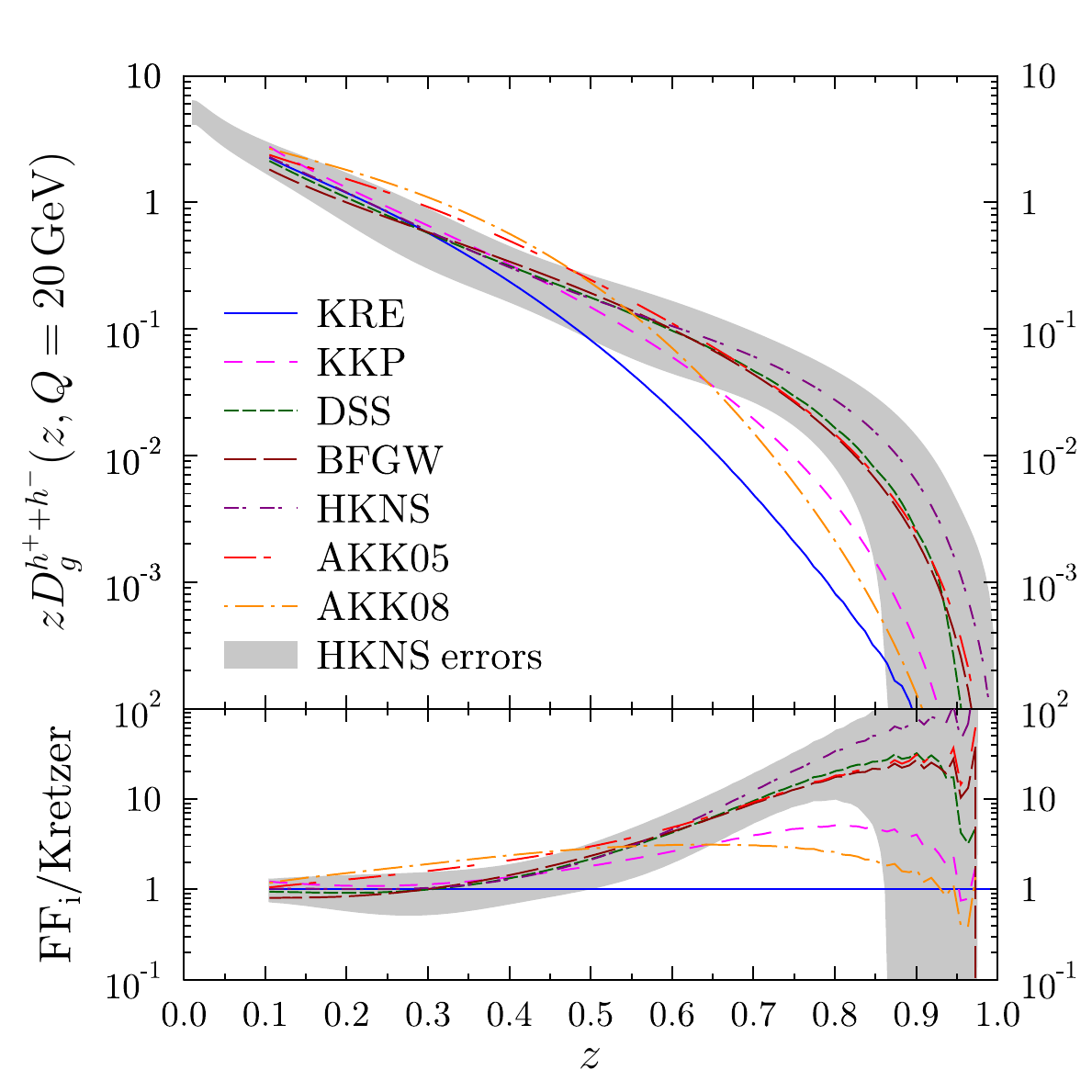}
\caption{Top: Charged-hadron fragmentation functions as a function of $z$ for $u$-quarks (left) and gluons
  (right) at $Q=20\,\rm{GeV}$. Bottom: Ratio between different FFs over the Kretzer FFs.} 
\label{fig:ff_comparison1}
\end{figure}

The main constraints in global fits of FFs come from the inclusive hadron production in $\epem$
collider experiments, from which the data are abundant~\cite{hepdata}. These experiments are, however, mainly
sensitive to the quark FFs leaving the gluon-to-hadron FFs largely unconstrained. Similarly, the data for semi-inclusive hadron
production in deeply inelastic scattering~\cite{Ashman:1991cj}, used in the \textsc{dss} fit, 
predominantly originates from the quark fragmentation. This leaves the gluon FFs prone to parametrization-bias
and other theoretical assumptions. To improve on this, \textsc{dss} and \textsc{akk08} include
various datasets (in different combinations) from hadronic collisions at RHIC~\cite{Adams:2006nd,Arsene:2004ux,Adler:2003pb}, 
Sp$\overline{\rm p}$S~\cite{Bocquet:1995jr,Albajar:1989an,Arnison:1982ed} and Tevatron~\cite{Abe:1988yu,Acosta:2005pk}.   
The bulk of these measurements is, however, concentrated at rather small values of transverse momentum
$\pT \lesssim~5~{\rm GeV/c}$ (dictating the hard scales of the process) where the sensitivity
to the gluon FFs is certainly present, but where the NLO pQCD calculations are not
well under control due to the large scale uncertainty (see later). Therefore, the extraction of the gluons
FFs based on these data cannot be considered completely safe, and, therefore, we believe that the ``older''
sets of FFs with only $\epem$ data should not be blindly discarded.
Although the uncertainties in FFs due to the experimental errors (or lack of data) have been addressed in some
FF extractions (see also Ref.~\cite{Epele:2012vg}), only \textsc{hkns} provides the necessary information to fully
propagate these uncertainties to further observables. The \textsc{bfgw} package provides three alternative sets
with somewhat different gluon FFs, but the mutual variation between these sets translates only to a few-percent
difference in the observables we discuss here. The error analysis of \textsc{dss} was performed via the method
of Lagrange multipliers \cite{Stump:2001gu} which does not allow the end user to estimate the propagation of
their FF errors. For these reasons,
in what follows, we will present the central results from all the parametrizations including also the \textsc{hkns} 
error bands. 

\begin{figure}[tbhp]
\centering
\includegraphics[width=0.49\textwidth]{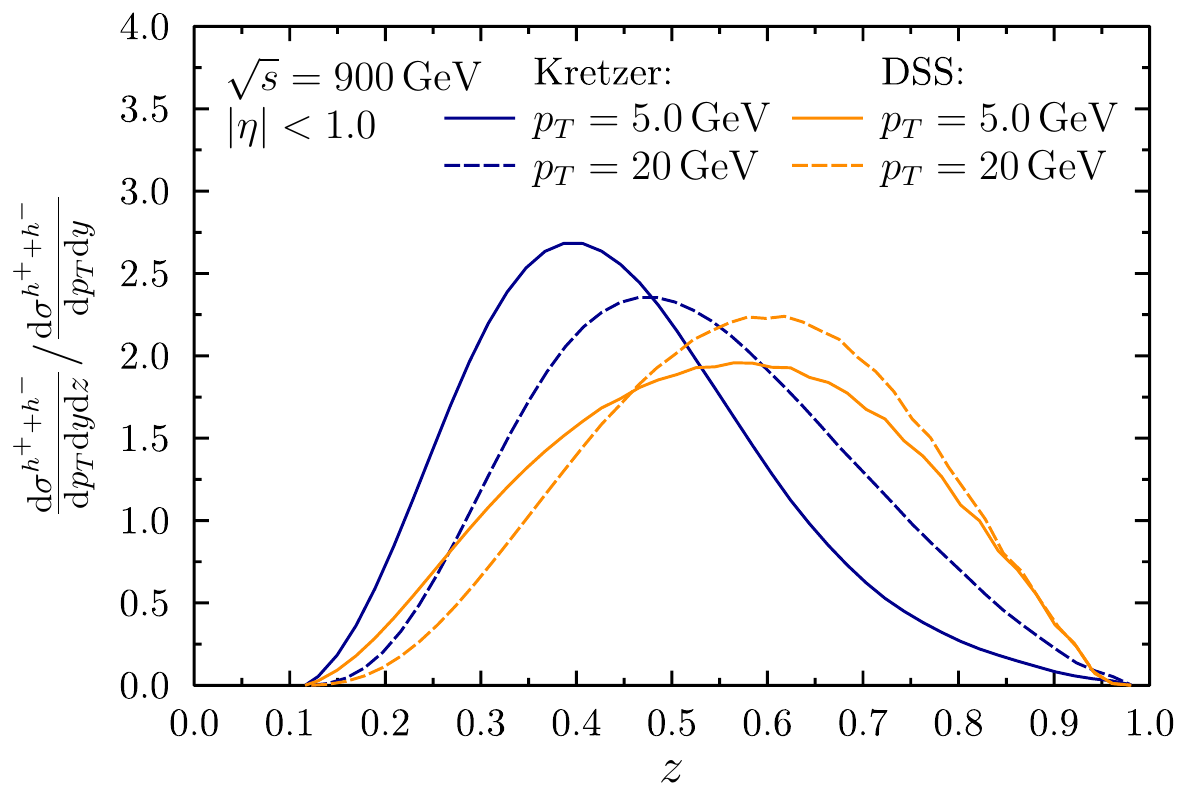}
\includegraphics[width=0.49\textwidth]{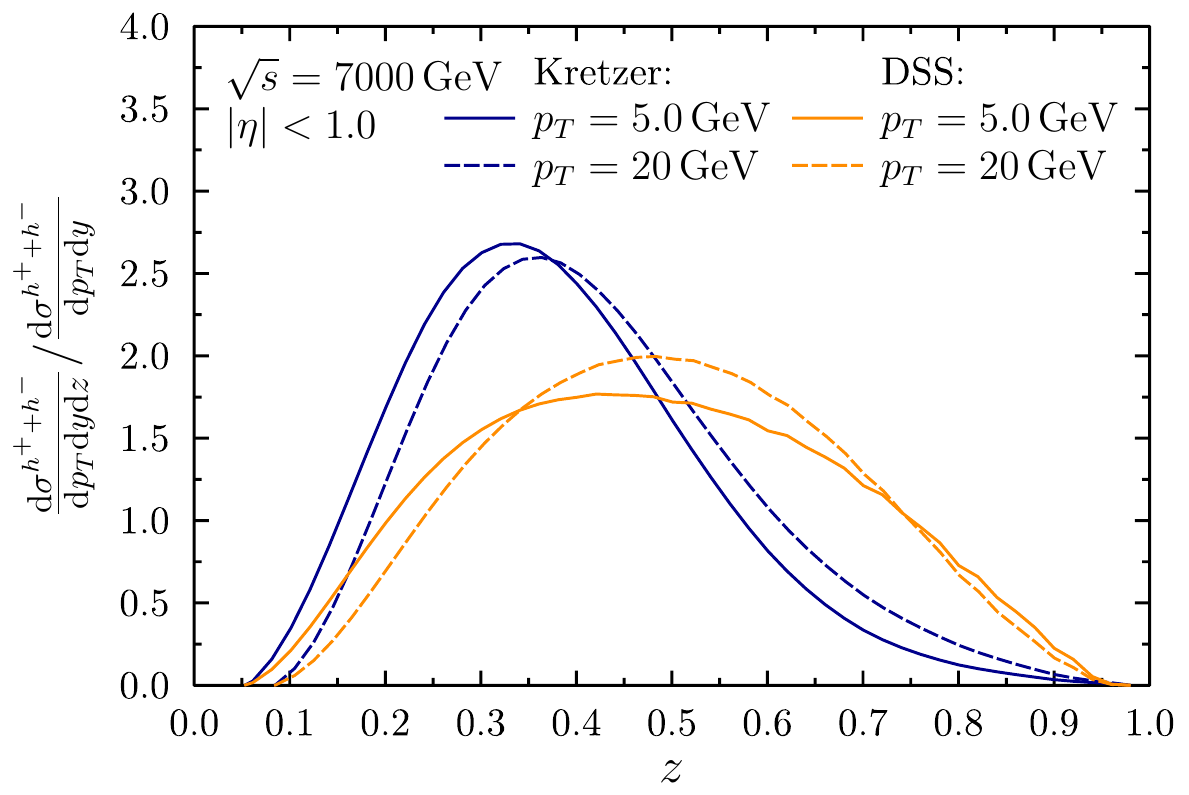}
\caption{Normalized cross section for charged-hadron production as a function of $z$ for
  $\sqrt{s}~=~900\,\rm{GeV}$ (left) and $\sqrt{s}=7000\,\rm{GeV}$ (right) for $\pT=5\,\rm{GeV/c}$ (solid)
  and $\pT=20\,\rm{GeV/c}$ (dashed) at midrapidity, obtained with Kretzer (dark blue) and \textsc{dss} (orange) FFs.} 
\label{fig:ff_comparison2}
\end{figure}

In Fig.~\ref{fig:ff_comparison1}, we present a comparison of the up-quark and gluon FFs into charged hadrons
for all the available FFs at a common scale $Q=20 \, {\rm GeV}$. 
The spread among the gluon FFs is
significantly larger than in the case of quarks --- a clear indication of lack of definitive constraints ---
in particular for moderate and large $z>$~0.3 values. 
Indeed, even the \textsc{hkns} error band is not broad enough to cover all different sets
above $z \sim 0.5$. We believe this is mainly due to the small amount of fit parameters that could be left
free in the absence of strict gluon constraints from the $\rm e^+\rm e^-$ data alone, as discussed above.
To understand at which $z$ values the hadron production at LHC energies probes the FFs, 
and to what extent this depends on the hardness of the gluon FFs, we examine
the shape of the differential $z$ distributions. A couple of typical examples 
corresponding to the LHC kinematics are shown in Fig.~\ref{fig:ff_comparison2}. 
The old Kretzer and the modern \textsc{dss} FFs are considered here
as the hardness of their gluon FFs is quite different.
The $z$ distributions appear to be rather broad and to depend significantly on the specific set of FFs used.
The spectra in the range $\pT$~=~5--20~GeV/c  probe average hadron fractional momenta $\mean{z}\approx$~0.4--0.6
at $\sqrts$~=~900~GeV, decreasing to $\mean{z}\approx$~0.3--0.5 at $\sqrts$~=~7~TeV. 
The behaviour of these $z$ distributions can be understood approximating the $\hat p_{3T}$ dependence of the
convolution between the PDFs and the partonic cross sections $d\hat\sigma$ by a power law, as follows~\cite{Eskola:2002kv}:
\begin{equation}
 \sum _{ij} \int dx_1 dx_2 f_i^{h_1}(x_1,\mu^2_{\rm fact}) f_j^{h_2}(x_2,\mu^2_{\rm fact})
 \, \frac{d\hat{\sigma}(\hat p_1^i + \hat p_2^j \rightarrow \hat p_3^\ell, \mu^2_{\rm ren}, \mu^2_{\rm fact}, \mu^2_{\rm frag})}{d\hat p_{3T} d\eta}
\approx C_\ell {\hat p_{3T}^{-n}},
\end{equation}
where $C_\ell$ is a $\pT$-independent constant, so that one can write
\begin{equation}
\frac{d\sigma(h_1 + h_2 \rightarrow h_3 + X)}{d\pT d\eta dz}  \approx \sum_\ell \frac{C_\ell}{z} \hat p_{3T}^{-n} D_{\ell \rightarrow h_3}(z,\mu^2_{\rm frag})
= \pT^{-n} \sum_\ell  C_\ell \, z^{n-1} D_{\ell \rightarrow h_3}(z,\mu^2_{\rm frag}).
\end{equation}
Due to the factor $z^{n-1}$, the contributions from small values of $z$ are efficiently
suppressed, and the average value of $z$ becomes much larger than the kinematic lower limit
$z_{\rm min} \approx 2\pT/\sqrt{s}$ at midrapidity. However, towards larger $\sqrt{s}$, the $\pT$
distributions are flatter (the power $n$ is smaller), and the $z$ distributions become
on average shifted towards smaller values of $z$.  In any case, the contributions from
the small-$z$ region, $z \lesssim 0.05 - 0.1$, which is more difficult to treat in DGLAP-based
approaches~\cite{deFlorian:1997zj,Sassot:2010bh}, remain very small and the use of the standard FF framework is
well justified.

\begin{figure}[tbhp]
\centering
\includegraphics[width=0.49\textwidth]{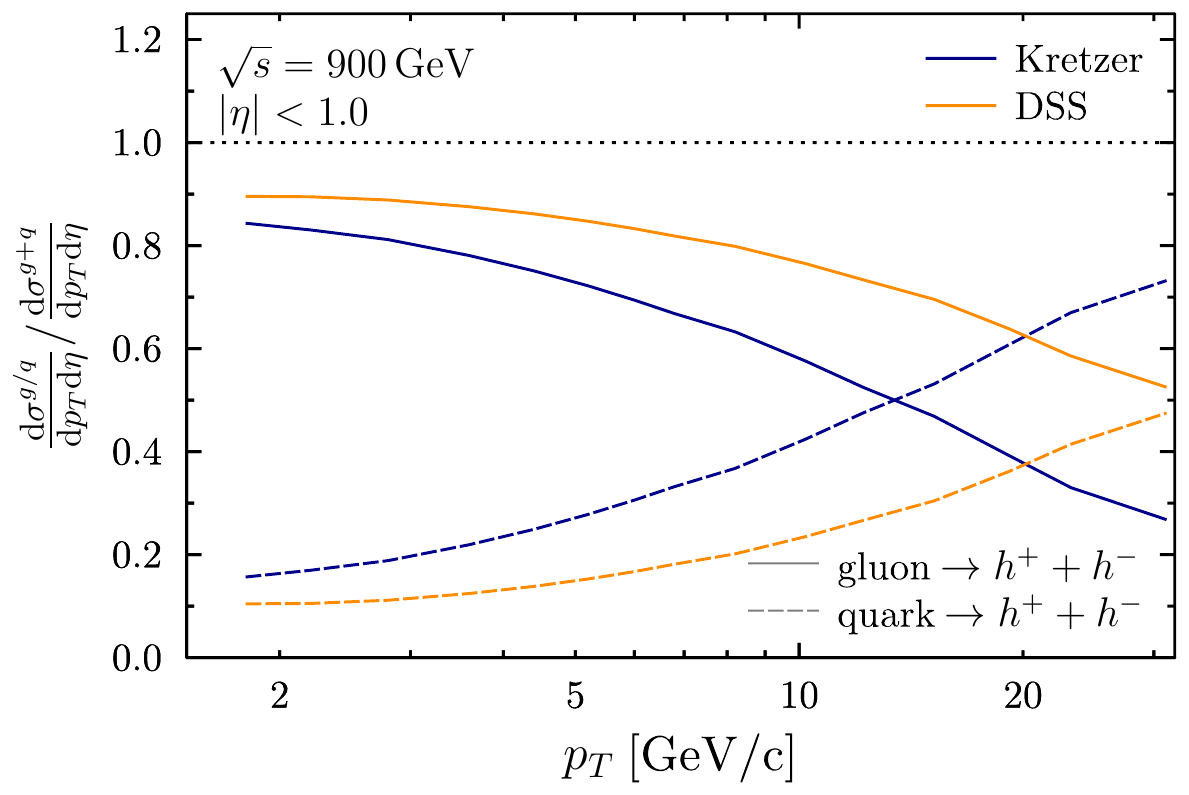}
\includegraphics[width=0.49\textwidth]{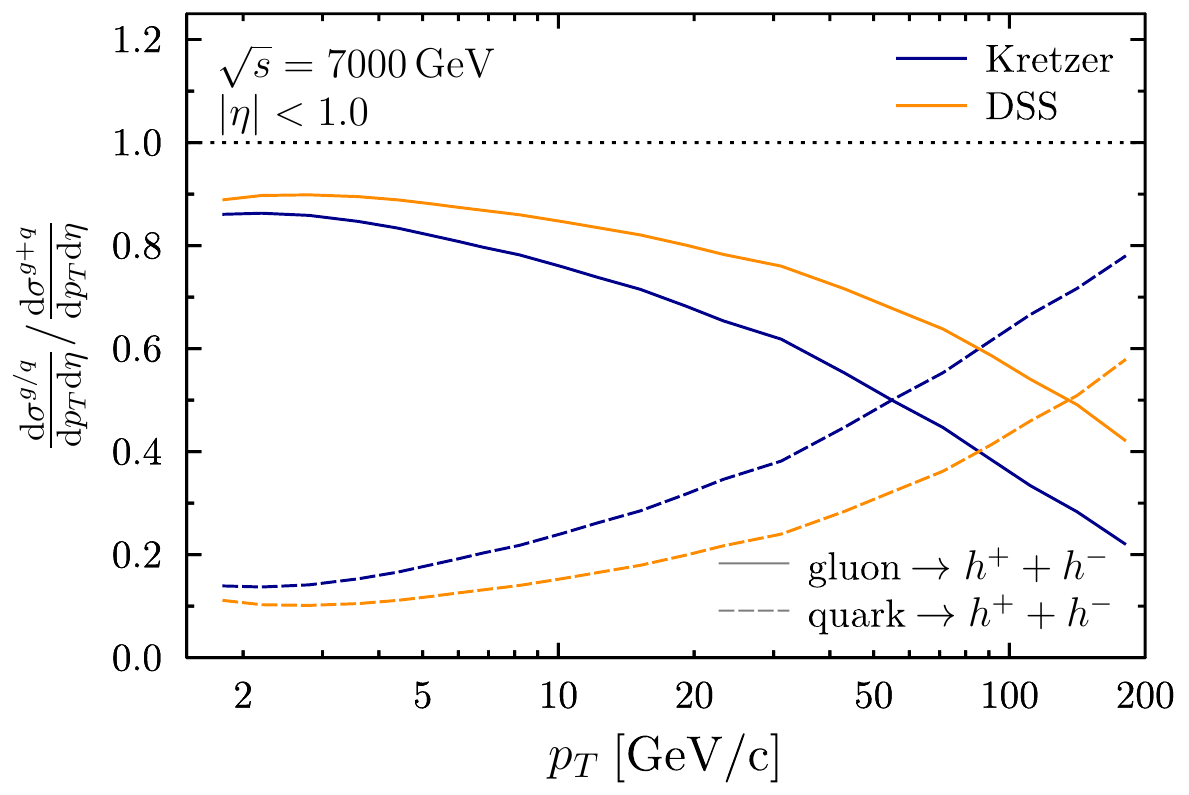}
\caption{Relative contributions of quark (dashed) and gluon (solid) fragmentation to the inclusive
  charged-hadron cross section at $\sqrt{s}=900\,\rm{GeV}$ (left) and $\sqrt{s}=7000\,\rm{GeV}$ (right) at
  midrapidity, obtained with Kretzer (dark blue) and \textsc{dss} (orange) FFs.} 
\label{fig:ff_comparison3}
\end{figure}

The relative contributions from the quark and gluon fragmentations are plotted in Fig.~\ref{fig:ff_comparison3}
for $\sqrt{s}=900\,\rm{GeV}$ (left) and $\sqrt{s}=7000\,\rm{GeV}$ (right).
At small $\pT$, the gluon fragmentation clearly dominates but towards large values of $\pT$ the quark fragmentation
becomes eventually predominant. In any case, the gluon contribution is always significant and therefore the LHC promises
to be a good ``laboratory'' to determine the gluon FFs in the region of $\pT$ where the NLO pQCD calculations can be
considered to be well under control.

\section{Comparison of NLO pQCD to high-$\pT$ charged-hadron collider data}

In this section, we compare the data from various experiments with the NLO calculations using the seven FF sets
listed in Table~\ref{tab:FFs}. Our main attention will be on the latest data from CMS~\cite{Chatrchyan:2011av,CMS:2012aa} and
ALICE~\cite{Abelev:2013ala} for \pp\ collisions at the LHC, as well as the CDF
measurements~\cite{Aaltonen:2009ne} in \ppbar\ collisions at Tevatron. We do not include the similar
ATLAS measurements~\cite{Aad:2010ac} here as their results have not been given as invariant cross
sections, but only in terms of the absolute yields. However, we have checked that the shapes of the ATLAS
$\pT$-spectra are in agreement with those measured by the other LHC experiments. In order to compare with some
earlier hadron-collider data, included in the \textsc{akk08} and \textsc{dss} global fits,
we consider also the \ppbar\ results from UA1~\cite{Bocquet:1995jr,Albajar:1989an,Arnison:1982ed} as well as the
\pp\ spectra measured by STAR~\cite{Adams:2006nd}.

\begin{figure}[tbhp]
\centering
\includegraphics[width=0.49\textwidth]{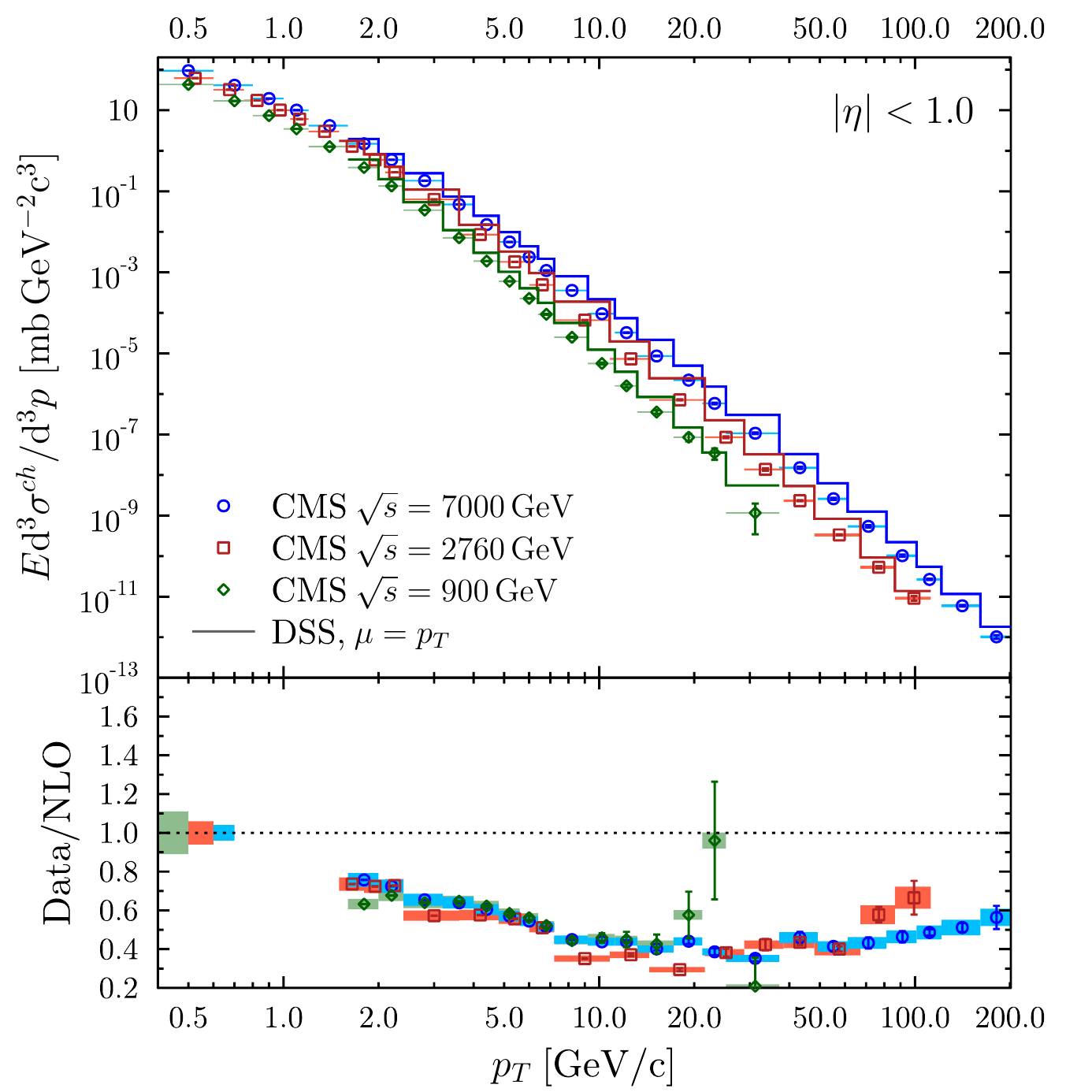}
\includegraphics[width=0.49\textwidth]{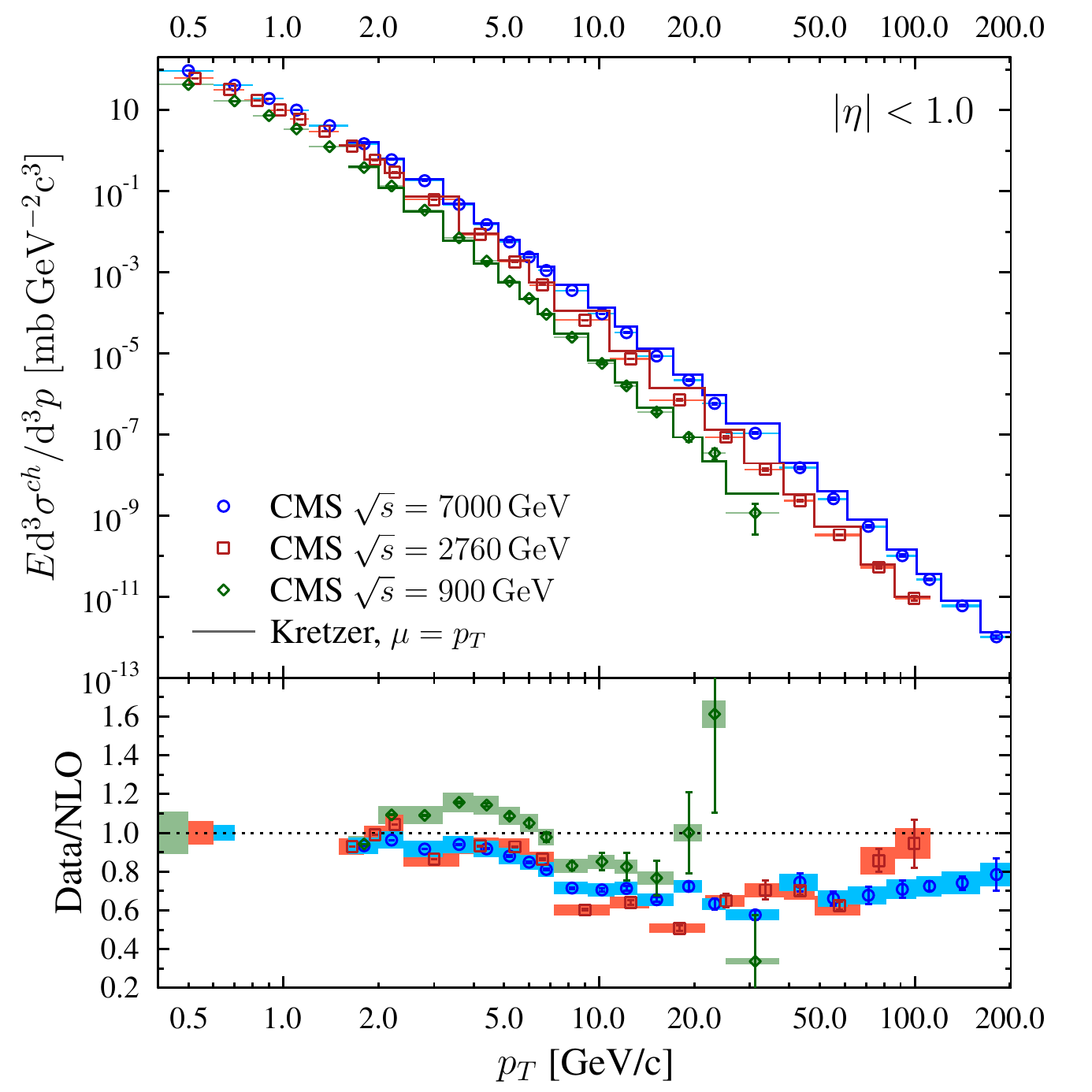}
\caption{Top: Charged-hadron invariant cross sections measured as a function of $\pT$ by
  CMS~\cite{Chatrchyan:2011av,CMS:2012aa} at $\sqrt{s}=0.9 \, {\rm TeV}$ (green diamonds),  
$\sqrt{s}=2.76 \, {\rm TeV}$ (red squares), and $\sqrt{s}=7 \, {\rm TeV}$ (blue circles),
compared to NLO calculations with \textsc{dss}~\cite{deFlorian:2007hc} (left) and Kretzer (right)~\cite{Kretzer:2000yf} FFs.
The point-to-point systematic and statistical errors are indicated by colored rectangles and error bars.
Bottom: Ratio between the data and the respective calculations. The boxes at the beginning of the $\pT$-axis
mark the luminosity uncertainties of each measurement.} 
\label{fig:AbsoluteSpectrum1}
\end{figure}

As an example of the $\pT$-differential cross sections, Fig.~\ref{fig:AbsoluteSpectrum1} presents
the CMS measurements for inclusive charged hadrons at $\sqrt{s}=0.9, 2.76, 7 \, {\rm TeV}$ at midrapidity.
The data, spanning five orders of magnitude in perturbatively-accessible values of $\pT$, 
are compared to the NLO calculations using two sets of FFs, \textsc{dss} and Kretzer.
While \textsc{dss} clearly overshoots the data (by up to a factor of 2), the Kretzer FFs do a much better
job in describing the spectra both in shape and absolute normalization. The apparent difference between the
two parametrizations derives from the fact that the large-$z$ Kretzer gluon FFs are
much softer than those of \textsc{dss} (Fig.~\ref{fig:ff_comparison1}).\\

\begin{figure}[tbh]
\centering
\includegraphics[width=\textwidth]{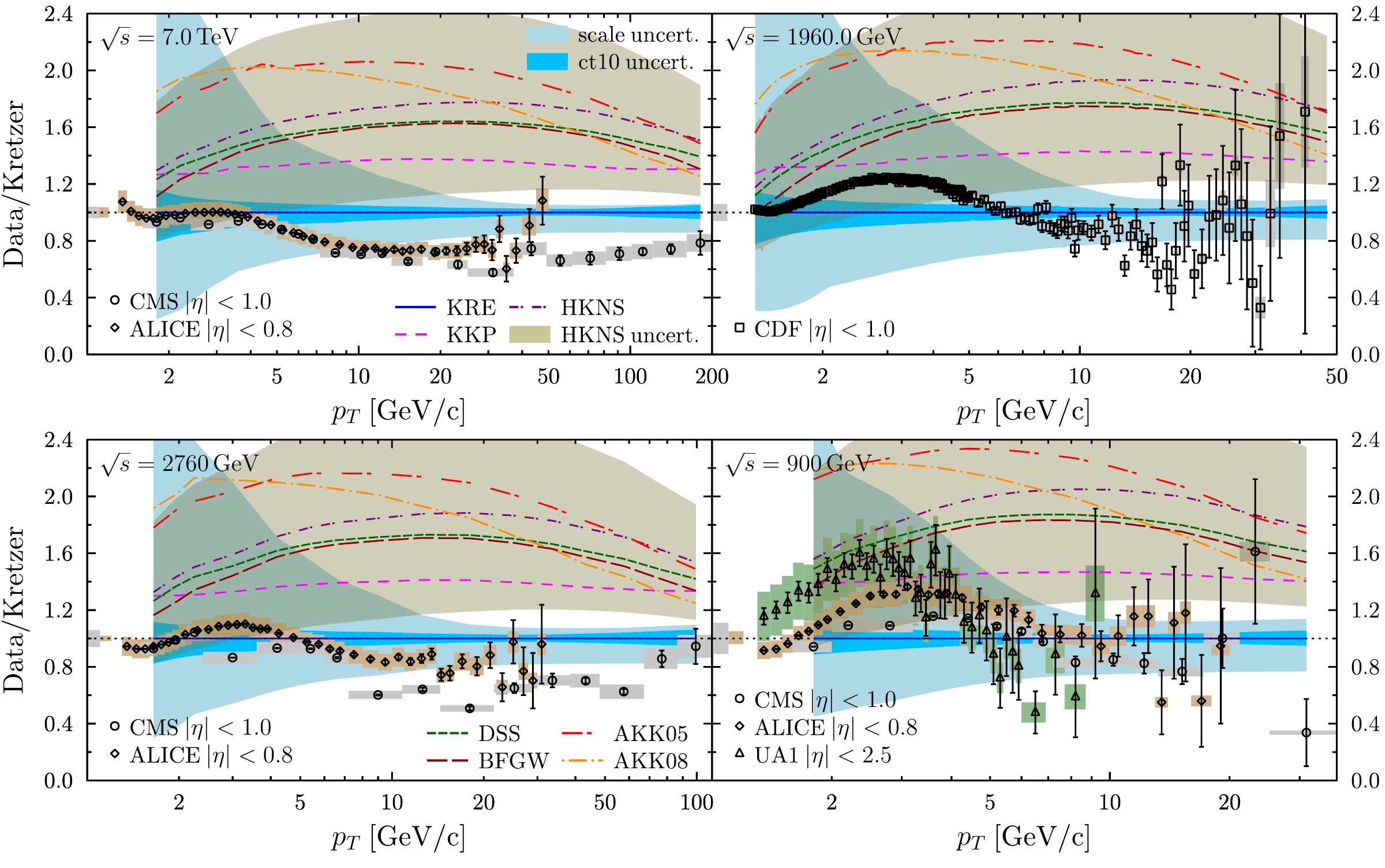}
\caption{Ratio of the inclusive charged-hadron spectra measured by CMS
  (circles)~\cite{Chatrchyan:2011av,CMS:2012aa}, ALICE~\cite{Abelev:2013ala} (diamonds), 
CDF (squares)~\cite{Aaltonen:2009ne}, and UA1 (triangles)~\cite{Albajar:1989an} at $\sqrts$~=~900--7000~GeV,
over the corresponding NLO calculations using the Kretzer FFs. The curves show the NLO predictions obtained with
other FF sets: \textsc{kkp} (pink scarcely dashed), \textsc{dss} (green dashed), \textsc{bfgw} (brown long-dashed), 
\textsc{hkns} (purple dashed-dotted), \textsc{akk08} (yellow dotted-dashed), and \textsc{akk05} (red long-dashed short-dashed)
relative to Kretzer FFs. The point-to-point systematic and statistical errors are indicated
by colored rectangles (gray for CMS and CDF, brown for ALICE, green for UA1) and error bars.
The boxes at the beginning of the $\pT$-axis mark the overall normalization uncertainty.
The light-blue bands correspond to the scale uncertainty envelopes while the dark-blue ones indicate
the variations derived from the \textsc{ct10nlo} PDF error sets. 
The \textsc{hkns} uncertainties are shown by the light-brown bands.
}
\label{fig:CompLHC}
\end{figure}

The central result of our paper is shown in Fig.~\ref{fig:CompLHC} which presents a comprehensive
comparison of the charged-hadron world-data in the TeV-range to the NLO predictions using the seven
most recent sets of FFs. The panels show the ratio of the data to the predictions obtained with the 
Kretzer FF (data points), as well as the ratios of cross sections obtained with various FFs over those from Kretzer
(curves). The light-blue band denotes the scale uncertainty envelope (cf. Eq.~(\ref{eq:scalevariation})), and the dark-blue band the uncertainty derived
from the \textsc{ct10nlo} PDF (90\% confidence level) error sets. 
The \textsc{hkns} error band is shown in light-brown color.
The scale uncertainty below 
$\pT\approx~10~\,{\rm GeV/c}$ is prohibitively large making it difficult to draw any strict conclusion
regarding the level of agreement between the data and the calculations there (although the central NLO
prediction obtained with the Kretzer FFs agrees very well with the data). 
The scale dependence, however, stabilizes below $\pm$20\% beyond $\pT \approx 10 \, {\rm GeV/c}$ and it 
is rather this region where the NLO calculations are to be fully trusted. In all cases, the PDF errors are
almost negligibly small in comparison to the scale uncertainty.\\

All the LHC data are in mutual agreement within their systematic and statistical uncertainties, although
especially the CMS data at 2.76~TeV seem to show larger fluctuations than those inferred from the typical size
of the quoted point-to-point experimental uncertainty. The UA1 spectrum at $\sqrt{s}=0.9 \, {\rm TeV}$ is
mostly above the CMS and ALICE results and is only barely compatible with them. However, as the
shape of the $\pT$ distribution is not incompatible with the rest, this disagreement could well
be an issue of the experimental determination of the overall normalization in the oldest measurement.\\

The results of Fig.~\ref{fig:CompLHC} exhibit clear systematic trends as a function of $\sqrt{s}$ and $\pT$:
As $\sqrt{s}$ increases from $0.9 \, {\rm TeV}$ to $7 \, {\rm TeV}$,
the experimental spectra gradually sink more and more below the theoretical predictions. However,
the shape of the data-to-theory ratio remains qualitatively similar regardless of
the collision energy. Especially, relative to the calculation with the Kretzer FFs, 
the data first show a ``bump'' at $\pT\approx$~4~GeV/c but then straighten up beyond $\pT\approx 10 \, {\rm GeV/c}$.
Indeed, the flatness of the data/theory ratio is worth noticing, although the absolute spectra span many orders of
magnitude. This suggests that the underlying pQCD dynamics of the hadron production is indeed correctly
understood and that the data-theory disagreement lies rather in the current sets of FFs.
On average, the Kretzer FFs used as reference for the data/NLO ratios shown in Fig.~\ref{fig:CompLHC},
seem to do  the best job in describing the data,
the results from all other FFs being practically enclosed by the \textsc{hkns} error bands.
We quantify the data-to-theory correspondence by computing the $\chi^2$ values for each FF set (in the case
of \textsc{hkns} only the central predictions are considered) defined by
\begin{equation}
 \chi^2 \equiv \sum_i \left( \frac{D_i-T_i}{\delta_i^{\rm tot}} \right)^{2},
\end{equation}
where $D_i$ corresponds to the data point with total error $\delta_i^{\rm tot}$
(correlated and uncorrelated point-to-point uncertainties added in quadrature), 
and the theory values $T_i$ are specific for each set of FFs. The sum runs
over all the data shown in Fig.~\ref{fig:CompLHC} using three different
cuts for the hadron transverse momenta: $\pT^{\rm min}~=~1.3, 5, 10 \, {\rm GeV/c}$.
Such thresholds are chosen so as to reduce the weight of the lower-$\pT$ data which would otherwise dominate the $\chi^2$ due
to their larger cross sections and associated smaller statistical uncertainties. The calculations are run 
for three scale choices,
$$
\mu \equiv (\mu_{\rm ren}/\pT,\mu_{\rm fact}/\pT,\mu_{\rm frag}/\pT) = \left({1 \over 2},{1 \over 2},{1 \over 2}\right), (1,1,1), (2,2,2)\,,
$$
which, above $\pT\approx 5 \, {\rm GeV/c}$, practically cover the larger selection of scale variations
used earlier, Eq.~(\ref{eq:scalevariation}).
The results from this exercise are shown in Table~\ref{tab:AAparams2}, and numerically confirm what is 
seen in Fig.~\ref{fig:CompLHC}: The lowest $\chi^2$ value is almost always
obtained with the Kretzer FFs and the highest one with \textsc{akk05}. The preferred choice
of scale is specific for each set of FFs and depends on the $\pT^{\rm min}$ cut imposed. However,
on average, the choice $\mu = (2,2,2)$ is preferred as this set of values tends to reduce the
computed cross sections and thereby moderate the data overshooting.

\begin{table*}[tbh]
\caption{Values of $\chi^2/N$ characterizing how the NLO calculations agree with the data at $\sqrt{s}~=~0.9-7$~TeV. Seven different FFs and three different scale-choices and minimum hadron-$\pT$ values have been considered. $N$ is the total number of the LHC, Tevatron and UA1 data points above the $\pT^{\rm min}$ cut.
In obtaining the $\chi^2$, all the data uncertainties have been added in quadrature.}
\begin{center}
\begin{tabular}{c|ccc|ccc|ccc}
\hline
$\mu$	 & \multicolumn{3}{c|}{(1,1,1)}	& \multicolumn{3}{c|}{(1/2,1/2,1/2)} & \multicolumn{3}{c}{(2,2,2)} \\
$\pT^{\rm min} \,\rm{[GeV/c]}$ & 1.3	& 5.0	 & 10.0	 	& 1.3	 & 5.0	 & 10.0	 	 & 1.3	 	 & 5.0	 	 & 10.0 \\
$N$	 & 368	 & 169	 & 103	 	& 368	 & 169	 & 103	 & 368	 & 169	 & 103 \\
\hline
\textsc{kre}	 & 5.512	 & 8.536	& 11.20	 & 32.94	 & 30.03	 & 23.06	 & 12.77	 & 4.034	 & 2.935 \\
\textsc{kkp}	 & 28.90	 & 51.63	& 56.81	 & 151.6	 & 143.3	 & 108.2	 & 9.216	 & 14.62	 & 19.25 \\
\textsc{dss}	 & 63.36	 & 112.3	& 114.2	 & 248.5	 & 319.5	 & 245.6	 & 16.68	 & 33.50	 & 40.96 \\
\textsc{hkns}	 & 85.80	 & 149.5	& 151.1	 & 303.9	 & 396.8	 & 312.6	 & 24.28	 & 48.59	 & 57.49 \\
\textsc{akk05}	 & 169.9	 & 236.7	& 218.4	 & 594.6	 & 619.0	 & 428.9	 & 51.39	 & 84.58	 & 89.88 \\
\textsc{akk08}	 & 150.1	 & 177.7	& 154.4	 & 566.6	 & 486.5	 & 300.3	 & 40.82	 & 59.13	 & 60.70 \\
\textsc{bfgw}	 & 57.15	 & 106.3	& 108.7	 & 203.2	 & 294.1	 & 230.5	 & 15.71	 & 31.65	 & 38.70 \\\hline
\end{tabular}
\end{center}
\label{tab:AAparams2}
\end{table*}

The same conclusion is reached if the $\chi^2$ is computed accounting separately for
the correlated systematic errors. In this particular case, the only known systematic parameter is
the overall normalization (given by all but UA1) and the $\chi^2$ can be expressed~\cite{Stump:2001gu} as
\begin{eqnarray}
 \chi^2_{\rm corr} & = & \sum_{\rm k \in data \, sets} \chi^2_{k} \\
 \chi^2_{k}         & = & \sum_i \left( \frac{f_k D_i-T_i}{\delta_i^{\rm uncor}} \right)^2 + \left (\frac{1-f_k}{\delta_i^{\rm norm}} \right)^2,
\label{eq:chicorr}
\end{eqnarray}
where $\delta_i^{\rm uncor}$ is the uncorrelated error, and $\delta_i^{\rm norm}$ the
quoted normalization error. The parameter $f_k$ is found by
minimizing the $\chi^2$. The corresponding results are listed in 
Table~\ref{tab:AAparams}. In comparison to the uncorrelated-uncertainties case, the $\chi^2$ 
values become generally somewhat lower, since, the calculated values are usually
quite above the data, and the improvement attained in the first
term in Eq.~(\ref{eq:chicorr}) exceeds the growth of the latter.
This procedure, however, often also leads to unnaturally large values of $|1-f_k|$
as the disagreement between the calculations and the data tends to be much beyond the
normalization uncertainty quoted by the experiments.
In any case, the values in Table~\ref{tab:AAparams} lead to the same conclusions as
those extracted from Table~\ref{tab:AAparams2} and the final outcome is the same regardless of
the way the $\chi^2$-function is defined, namely that NLO predictions generally overpredict the experimental
charged-hadron spectra by a factor of two. Similar discrepancies have been found for high-$\pT$ neutral pion
and $\eta$ meson production at LHC energies~\cite{Abelev:2012cn} 
implying that the problem is not
limited to the total ($g\to h^{+}+h^{-}$) fragmentation function but affects the identified
($\pi^\pm$, $K^\pm$, $p$, $\overline{p}$)
gluon FFs individually.

\begin{table*}[tbh]
\caption{As Table~\ref{tab:AAparams2}, but accounting for the normalization uncertainty
of the data as in Eq.~(\ref{eq:chicorr}).}
\begin{center}
\begin{tabular}{c|ccc|ccc|ccc}
\hline
$\mu$	 & \multicolumn{3}{c|}{(1,1,1)}	& \multicolumn{3}{c|}{(1/2,1/2,1/2)} & \multicolumn{3}{c}{(2,2,2)} \\
$\pT^{\rm min} \,\rm{[GeV/c]}$ & 1.3	& 5.0	 & 10.0	 	& 1.3	 & 5.0	 & 10.0	 	 & 1.3	 	 & 5.0	 	 & 10.0 \\
$N$	 & 368	 & 169	 & 103	 	& 368	 & 169	 & 103	 & 368	 & 169	 & 103 \\
\hline
\textsc{kre}	 & 5.460	 & 5.063	 & 5.158	 & 11.73	 & 9.675	 & 10.25	 & 5.020	 & 3.402	 & 2.031 \\
\textsc{kkp}	 & 14.54	 & 17.97	 & 22.34	 & 36.06	 & 37.69	 & 43.16	 & 8.893	 & 7.648	 & 8.061 \\
\textsc{dss}	 & 31.74	 & 35.11	 & 44.75	 & 113.4	 & 82.19	 & 98.73	 & 13.86	 & 14.00	 & 16.59 \\
\textsc{hkns}	 & 41.30	 & 45.73	 & 58.50	 & 130.9	 & 101.1	 & 123.9	 & 17.58	 & 18.67	 & 22.72 \\
\textsc{akk05}	 & 44.89	 & 66.01	 & 86.08	 & 208.4	 & 159.0	 & 175.9	 & 21.00	 & 27.13	 & 35.42 \\
\textsc{akk08}	 & 29.72	 & 49.73	 & 63.47	 & 160.3	 & 137.1	 & 130.9	 & 15.79	 & 19.42	 & 25.51 \\
\textsc{bfgw}	 & 35.61	 & 33.91	 & 43.09	 & 118.0	 & 77.23	 & 93.59	 & 13.86	 & 13.61	 & 16.18 \\\hline
\end{tabular}
\end{center}
\label{tab:AAparams}
\end{table*}

\begin{figure}[tbhp]
\centering
\includegraphics[width=\textwidth]{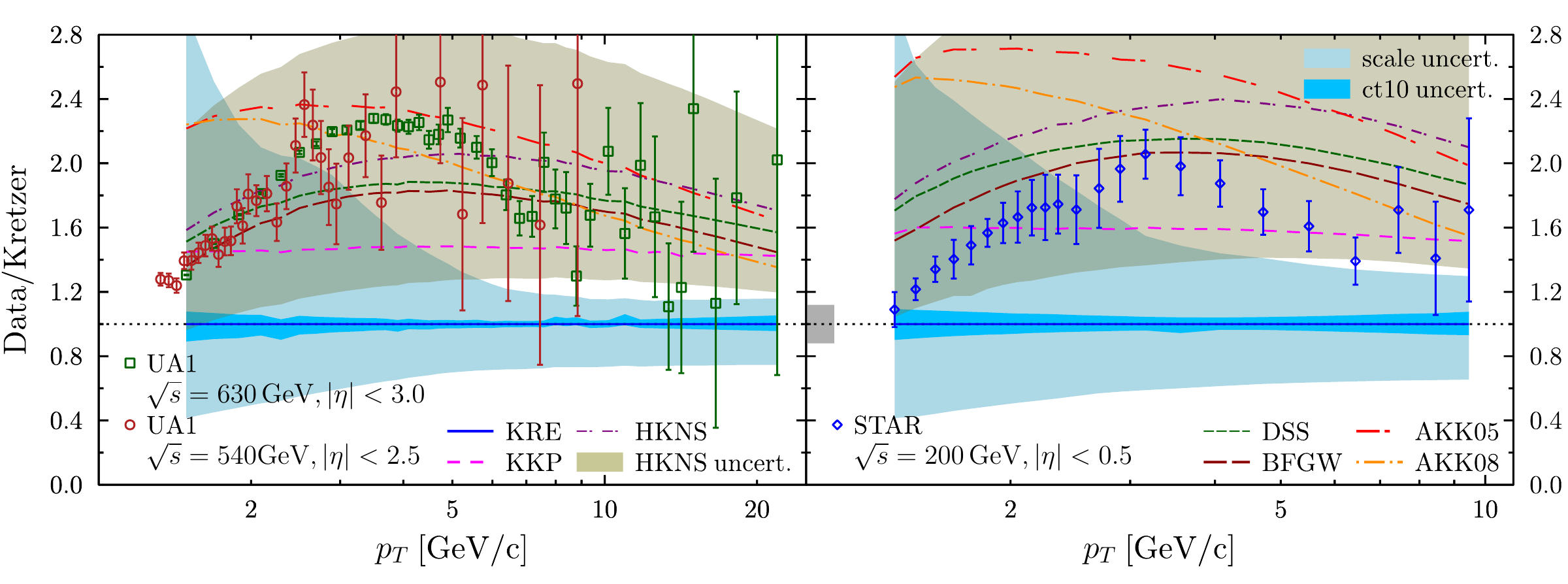}
\caption{Ratios data/NLO[\textsc{kre}] (data points) and NLO[FFs]/NLO[\textsc{kre}] (curves) as in
  Fig.~\ref{fig:CompLHC} but for lower-energy  UA1 ($\sqrts$~=~540,630~GeV, left panel)
  \cite{Bocquet:1995jr,Albajar:1989an,Arnison:1982ed} and STAR ($\sqrts$~=~200~GeV, right
  panel)~\cite{Adams:2006nd} \pp\ and \ppbar\ collisions.}
\label{fig:CompOld}
\end{figure}

Finally, we take a look at collider data at lower c.m. energies, $\sqrt{s}$~=~200--630~GeV, where the situation 
is different than that found at the LHC energies. Fig.~\ref{fig:CompOld}
presents a comparison of NLO calculations to the UA1~\cite{Bocquet:1995jr,Albajar:1989an,Arnison:1982ed}
and STAR~\cite{Adams:2006nd} charged-hadron spectra. For these datasets, the calculation obtained with the
Kretzer FFs --- the preferred set at the Tevatron and LHC energies --- gives a bad description of
the spectra. 
On the other hand, the \textsc{dss} set describes now these measurements reasonably well.
This is, however, not too surprising as these datasets were included in their actual fit.
In this case, also the \textsc{hkns} error band is wide enough to enclose these data.

Looking back to Fig.~\ref{fig:ff_comparison1}, one can see that the lower-energy collisions prefer much
harder gluons at large $z$ than the LHC data. That is, any set of FFs that can, more or less, reproduce
the lower-$\sqrt{s}$ data (preferring hard gluon FFs), will disastrously overshoot the LHC measurements
(preferring softer gluon FFs). 
As the variation in the probed range in $z$ is only mild as a function of $\sqrt{s}$ and $\pT$
(see Fig.~\ref{fig:ff_comparison2}) such a result hints that it may be difficult to 
tensionlessly include all these existing data into a global FF fit with a $\pT$ cut as low as e.g. 
$\pT \geq 2 \, {\rm GeV/c}$.
Indeed, as the very large scale-uncertainty indicates, the fixed-order NLO calculations are 
not stable below $\pT \approx 10 \, {\rm GeV/c}$, and it is questionable whether these lower-$\pT$ data points
should be even considered in such a fit in the absence of full NNLO corrections. 
It should be also recalled that within the RHIC kinematics reach, 
the threshold logarithms can still play a role by increasing the NLO cross sections for increasingly-high $\pT$
values, although such effects should die out towards larger $\sqrt{s}$ (at fixed $\pT$). In any case, the
fact that such effects cannot be seen in Fig.~\ref{fig:CompOld}, signals that threshold resummations are
likely not the main cause for the different $\sqrt{s}$-dependence of the NLO calculations and the data.\\

In addition, the good agreement between fixed-order NLO calculations~\cite{Wobisch:2011ij,Nagy:2003tz} 
and the jet data at LHC~\cite{Chatrchyan:2012bja,Aad:2011fc},
Tevatron~\cite{Aaltonen:2008eq,Abazov:2008hua} and RHIC~\cite{Abelev:2006uq}, and the fact that the single
high-$\pT$ charged particle spectrum is dominated by leading hadrons carrying out a large fraction,
$\mean{z}\approx$~0.5, of the parent parton (jet) energy (Fig.~\ref{fig:ff_comparison2}), strongly reinforces
the idea that the origin of the data-theory disagreement lies in an imperfect knowledge of the final
gluon-to-hadron fragmentation functions.\\

As a matter of fact, at low values of $\pT$ the whole picture of independent parton-to-hadron fragmentation
may not be adequate, especially in the case of production of heavier baryons. Higher-twist effects, 
where the hadron is produced directly (i.e. more exclusively) in the hard subprocess rather than by gluon or quark jet fragmentation,
may contribute to the cross sections at RHIC energies in the range of transverse momenta experimentally
studied~\cite{Arleo:2009ch} and hence ``contaminate'' the extraction of FFs in global fits that use such data.
Even at LHC energies, the proton-to-pion ratio below $\pT\approx 6 \, {\rm GeV/c}$
(see e.g.~\cite{RobertoPreghenellafortheALICE:2013yua}) appears to behave qualitatively very differently than
the kaon-to-pion ratio or the pQCD expectations. While the kaon-to-pion ratio increases smoothly towards
larger $\pT$, the proton-to-pion ratio contains a clear ``bump`` around $\pT\approx 3 \, {\rm GeV/c}$.
To reproduce such a behaviour, additional effects outside the pQCD toolbox are called for.
Assuming that the behaviour of the perturbative proton-to-pion ratio is qualitatively
similar to the the kaon-to-pion ratio, one could crudely estimate that there is a roughly $5\%$ 
``non-fragmentation'' enhancement around $\pT\approx 3 \, {\rm GeV/c}$ which then diminishes towards
higher $\pT$. Subtracting such a non-perturbative contribution from the LHC data would make
the disagreement between the data and e.g. \textsc{dss} even worse. On the contrary, in comparison to the
description with the Kretzer FFs, the data-to-theory ratio would be flatter and thereby
improve the compatibility. For lower $\sqrt{s}$ the surplus of (anti)protons
could be even more pronounced and remain important up to higher values of $\pT$. This could 
partly explain the strong $\sqrt{s}$-dependence of the data-to-theory ratio.
In Ref.~\cite{Abelev:2013ala}, it was observed that the ratios of the ALICE cross sections between
different but nearby $\sqrt{s}$ become rather well reproduced --- also at low $\pT$ ---  by the \textsc{dss}
FFs. However, it is important to note that as the $\sqrt{s}$ dependence of the $z$ distributions
(see Fig.~\ref{fig:ff_comparison2}) is only mild, part of the FF dependence is bound to cancel in such ratios.
Despite such cancellations, it appears that some $\sqrt{s}$ dependence still remains at low $\pT$, supporting
our conjecture regarding the presence of a $\sqrt{s}$-dependent non-perturbative component.

\section{Summary and outlook}
                
We have examined the LHC, Tevatron, Sp$\overline{\rm p}$S and RHIC data for inclusive unidentified
charged-hadron production at $\sqrt{s}=0.2 - 7 \, {\rm TeV}$ against NLO pQCD calculations with seven 
different sets of FFs, quantifying also the systematics associated with the scale and PDF uncertainties.
 The spread among the predictions with different FFs is
large and can be traced back to sizable mutual differences in the gluon-to-hadron FFs.
None of the existing FF sets can reproduce the experimental results optimally, on average,
clearly overshooting the data by up to a factor of two in the large-$\pT$ region, $\pT \geq 10 \, {\rm GeV/c}$,
where the fixed-order NLO pQCD calculations 
should be trustworthy as their scale dependence is modest.
The best overall agreement with the data is 
obtained using the relatively old Kretzer FFs in which the gluon-to-hadron FFs are the softest of all.
Below $\pT\approx 10 \, {\rm GeV/c}$, the $\sqrt{s}$-dependence of the data appears too strong
to be reproduced by calculations based solely on collinear factorization (especially
if also lower values of $\sqrt{s}$ are included in the comparison), although the NLO scale uncertainties
become there very large preventing one from drawing definitive conclusions. 
However, this may not be a problem of the pQCD calculation itself
as below $\pT\approx 10 \, {\rm GeV/c}$ there are increasing indications of 
additional non-perturbative effects in the case of (anti)proton production
even at the LHC energies, which may have a non-negligible impact on the total hadron yield.
These observations indicate that only the region above $\pT\approx  10 \, {\rm GeV/c}$ of these charged-hadron
data, with theoretical scale uncertainties below $\pm$20\%, should be included in forthcoming global
fits of parton-to-hadron fragmentation functions.   

\section*{Acknowledgments}
\noindent We acknowledge the financial support from the Magnus Ehrnrooth Foundation (I.H.) and from the
Academy of Finland, Project No. 133005.

\end{document}